\documentclass{aastex}
\usepackage{spr-astr-addons}
\usepackage{url}

\RequirePackage{color}

\begin{document}

\title{Galaxy rotation curves: the effect of $\vec j \times \vec B$ force}
\slugcomment{Not to appear in Nonlearned J., 45.}
%% Running heads
\shorttitle{Galaxy rotation curves}
\shortauthors{Tsiklauri}

\author{D. Tsiklauri} 
\affil{Astronomy Unit, School of Mathematical Sciences,
Queen Mary, University of London,
Mile End Road, London, E1 4NS, United Kingdom.}
%\email{\emaila}

\begin{abstract}
Using the Galaxy as an example, we study the effect of $\vec j \times \vec B$ force
on the rotational curves of gas and plasma in 
galaxies. 
Acceptable model for the galactic magnetic field and 
plausible physical parameters are used to fit the  
 flat rotational curve for gas and plasma 
based on the observed baryonic (visible) matter distribution and
$\vec j \times \vec B$ force term in the static MHD equation of motion.
We also study the effects of varied 
strength of the magnetic field, its pitch angle and length scale 
on the rotational curves.
We show that $\vec j \times \vec B$ force does not play an important role on the plasma dynamics
in the intermediate range of distances $6-12$ kpc from the centre, whilst the effect is sizable for 
larger $r$ ($r \geq 15$ kpc), where it is the most crucial. 
\end{abstract}

\section{Introduction}

Observed flat rotational curves of many galaxies have been the subject of
long-term controversy. The observational fact that the azimuthal
velocity of gas and stars in the galactic plane is constant over a large
range of the distances from the centre of a galaxy has yielded
two main explanations. In an attempt to save the assertion that the 
Newtonian gravitational theory holds over the
cosmological distances, one such theory assumes
the presence of non-baryonic massive dark halo surrounding
a spiral disk. In this scenario, gravitational acceleration $GM_<(r)/r^2$ which 
balances the centrifugal acceleration $V^2(r)/r$, is assumed to vary as $1/r$. 
This means that the mass enclosed within a certain radius $r$, $M_<(r)$, scales
as $\propto r$. However, this is not what is observed at large radii of the Galaxy. 
The second possible explanation of the flat rotational curves is that the 
Newtonian gravity does not apply on cosmological scales and further 
modifications are due (Milgrom 1983). Historically the latter explanation was 
not favoured due to absence of the general relativistic extension of the theory. 
However, this drawback was alleviated by the formulation of the
generalisation of Einstein's general relativity based on a pseudo-Riemannian 
metric tensor and a skew-symmetric rank three tensor field, called metric-skew-tensor gravity (MSTG). 
The latter leads to a modified acceleration law that can explain the flat rotation curves of 
galaxies and cluster lensing without postulating exotic dark matter (Moffat 2005). 
Recently, Brownstein \& Moffat (2006) have shown that MSTG can provide a good explanation to the 
flat rotational curves of a large sample of low and high surface brightness 
galaxies and an elliptical galaxy. Their MSTG fits were compared to those obtained 
using Milgrom's phenomenological MOND model and to the predictions of the Newtonian / Kepler acceleration law. 

In this work, using the Galaxy as an example, we study the effect of $\vec j \times \vec B$ force
on the rotational curves of gas and plasma in 
galaxies. We use an acceptable model for the galactic magnetic field and 
plausible physical parameters to fit the  
flat rotational curve for gas and plasma 
based on the observed baryonic (visible) matter distribution and
$\vec j \times \vec B$ force term in the static MHD equation of motion.
It should be mentioned that the present 
work was complete when author became aware of a similar earlier work by Nelson (1988).
The latter studied the dynamical effect of magnetic stress on the tenuous outer gaseous discs of galaxies.
Nelson (1988) used an earlier, less observationally constrained model (Sofue et al. 1986) for the magnetic field.
No fit to an observational rotational galactic curve was presented.
Here we advance the earlier hypothesis by choosing a more realistic model for the
galactic magnetic field, as well as perform fit to the Milky Way rotational curve.

Other significant previous developments include:  Battaner et al.~(1992) who
argue that an azimuthal magnetic field can carry slightly ionised gas 
with the general galactic rotation, rendering dark matter unnecessary. It was shown 
for the illustrative case of M31, that a magnetic field of 6 $\mu$G is required, and 
the synchrotron emission of relativistic electrons in this field is compatible 
with the observations. However this was not without a subsequent debate (Katz 1994).
More recent arguments in favour of magnetic fields, in this context, were also presented
in (Battaner \& Florido 2000; Battaner et al. 2002; Battaner \& Florido 2007). However, based on virial constraints
Sanchez-Salcedo \& Reyes-Ruiz (2004), on contrary, show that azimuthal magnetic fields hardly speed up H I disks of galaxies as a whole.
This demonstrates that the role of magnetic fields in the rotational curves of galaxies
is an active area of research  and there is no general agreement to date.

Our analysis shows that $\vec j \times \vec B$ force does not play an important role on the plasma and gas dynamics
in the intermediate range of distances $6-12$ kpc from the centre, whilst the effect of the force 
is significant for larger $r$ ($r>15$ kpc).  

\section{The model}

Let us consider the balance of forces acting on a small volume element of a galaxy. 
This is given by MHD equation of motion, which in its 
static ($\partial  / \partial  t =0$),  form reads as (e.g. Gosling \& Pizzo (1999)):
\begin{equation}
(\vec V \nabla ) \vec V + 2 \vec \omega \times \vec V +  
\vec \omega \times (\vec \omega \times \vec r) = - 
\frac{G M_<(r)}{r^2} \hat r + \frac{\vec j \times \vec B}{\rho(r)}.
%1
\end{equation}
The latter is written in the frame rotating with the galaxy, which does 
so with uniform angular momentum $\vec \omega \parallel$ to $z$. 
$\hat r$ is a unit vector along radial coordinate, $r$. We use 
cylindrical coordinate system $(r,\phi,z)$.
As it is common in the MHD we ignore the displacement current and 
assume $\vec j = \nabla \times \vec B / \mu_0$. Note that we ignore 
the gas pressure in Eq.(1) as a simplifying assumption. 

In the simplest possible case, whilst retaining essential physics, 
we assume that the radial velocity of the galactic matter is zero,  
$V_r=0$, and the only component of the velocity is azimuthal.  Thus, 
we end up with the only $r$-component of the MHD equation of motion Eq.(1):
\begin{equation}
-\frac{V_{\phi}^2(r)}{r} -2 \omega V_{\phi}(r) - \omega^2 r = 
- \frac{G M_<(r)}{r^2} + \frac{(\nabla \times \vec B \times \vec B )_r}{\mu_0 \rho(r)}.
%2
\end{equation}
The latter quadratic equation can be solved to yield 
\begin{equation}
V_{\phi}(r)= -\omega r \pm \sqrt{  \frac{G M_<(r)}{r} - \frac{r (\nabla \times \vec B \times \vec B )_r }{\mu_0 \rho(r)}}.
%3
\end{equation}
Here $V_{\phi}(r)$ is the azimuthal velocity in the rotating frame. It is related to the rotational 
(azimuthal) velocity in the laboratory frame, $\tilde V_{\phi}(r)$, via
\begin{equation}
\tilde V_{\phi}(r)= V_{\phi}(r) + \omega r = 
\pm \sqrt{  \frac{G M_<(r)}{r} - 
\frac{r (\nabla \times \vec B \times \vec B )_r }{\mu_0 \rho(r)}}. 
%4
\end{equation}
As expected, we could have arrived at the same result by writing the MHD equation of motion 
in the non-inertial frame rotating with the Galaxy as MOND and MSTG models do. 
Note that $\pm$ signs refer to the galactic rotational curves for the either 
side of the galactic centre on a given azimuth direction.

Naturally, one could perform a study that would include many galaxies. 
However, our aim here is to demonstrate the principle. Thus, our case 
study will be the Galaxy and in what follows we fix 
$|\vec \omega| = \omega$ at $\omega= 2 \pi/(250 \times 10^6\times 365.25\times 24 \times 60 \times 60)$ rad s$^{-1}$, 
as we know that the Galaxy rotates once in 250 million years.

As far as the distribution of ordinary, baryonic matter is concerned, we use 
observationally constrained model presented in (Brownstein \& Moffat 2006). In particular, 
they used a simple model for $M(r)$ 
\begin{equation}
M(r)= M\left(\frac{r}{r_c+r}\right)^3,
%5
\end{equation}
with the best fit parameters for the galaxy being $M=9.12\times10^{10}$ $M_{\sun}$ and $ r_c = 1.04$ kpc. 
The density prescribed by the same model (Brownstein \& Moffat 2006), is 
\begin{equation}
\rho(r)=   \frac{3}{4 \pi r^3} M(r) \left(\frac{r_c}{r_c + r}\right).
%6
\end{equation}

The key ingredient of our model is the magnetic field, which manifests itself in 
the rotational curve via additional force 
${\vec j \times \vec B}/ {\rho(r)}= {\nabla \times \vec B \times \vec B} /({\rho(r)} \mu_0 )$ 
in Eq.(4). There is a body of work that is dedicated to the determination of galactic 
magnetic fields (see for a review Vall\'ee (2004)) including the Galaxy (Han \& Qiao 1994) or to 
modelling these by the numerical simulations (Dobbs \& Price 2008). The galactic magnetic field 
models are commonly used in the study of cosmic rays (Stanev 1997; Alvarez-Muniz et al.~2002), due to 
the important ability of magnetic component of the Lorentz force to alter or trap 
charged cosmic ray particle paths. We, on contrary,  use the galactic magnetic 
field here in the context of the galactic rotational curves.
There are several methods that enable to infer galactic magnetic fields. These include: 
studies of starlight polarisation, background radio emission, Zeeman splitting, 
the rotational and dispersion measures of pulsars and extragalactic radio 
sources (Han \& Qiao 1994). The main outcome of these studies is that the magnetic field of the 
galaxy has Bi-Symmetric Spiral (BSS) configuration, which is given by 
\begin{eqnarray}
B_{\phi}=B_0(r)\cos (\phi - \beta \ln(r/r_0)) \cos p, \\
B_{r}=B_0(r)\cos (\phi - \beta \ln(r/r_0)) \sin p, \\
B_z=0.
%7-9
\end{eqnarray}
Here $\beta= 1 / \tan(p)$.
Han \& Qiao (1994) used Eqs.(7-9) to fit them to the rotation measures of pulsars and extragalactic
courses. They used $B_0(r)=const$ approximation for their fit, and found the following best fit
parameters: $B_0=(1.8 \pm 0.3)$ $\mu$G, the pitch angle $ p=-8.2^\circ \pm 0.5^\circ$, and
$r_0=(11.9 \pm 0.15)$ kpc.

Note that taking into account Eqs.(7-9), Eq.(4) can be rewritten as
\begin{equation}
\tilde V_{\phi}(r)=  
\pm \sqrt{  \frac{G M_<(r)}{r} + 
\left( \frac{B_\phi}{r} +\frac{\partial B_\phi}{\partial r} - 
\frac{1}{r}\frac{\partial B_r}{\partial \phi} \right)\frac{r B_\phi}{\mu_0 \rho(r)}
}. 
%10
\end{equation}

\section{results}

We performed  fit of our analytical model (Eq.(10)) to the observational 
rotational curve of the Galaxy (see Fig.(3) from (Brownstein \& Moffat 2006)) using
non-linear least squares Marquardt-Levenberg algorithm.
A set of vastly different "initial guess" parameters were used.
Generally, convergence of the fit without fixing $B_0$
was not possible. Thus to achieve the fit convergence
it was necessary to fix $B_0$ and vary $p$ and $r_0$ 
during the fit.
We stopped our choice on value $B_0= 6.8 $ $\mu$G, because it provides
such fit that the best fit curve is barely consistent with lower tip
of the data point error bar at $r=15$ kpc in Fig.1. This provides
a tolerable fit of the model to the data in the range $r \geq 15$ kpc.
Thus, the fit with fixed $B_0= 6.8 $ $\mu$G yields:
$p= (24.12 \pm 1.27)^\circ$, $r_0= (4.54 \pm 0.37)$ kpc. 
Note that, naturally, the results for $p= -24.12^\circ$ are identical
to $p= 24.12^\circ$.
Thus, we regard these values as the best fit parameters of our model. The rotational curve with the best fit
parameters is shown with a thin sold line in Figs.(1)-(4).
Note that solar system position is at $\phi=0$ (fixed value used in this work).

\begin{figure}
\includegraphics[width=\columnwidth]{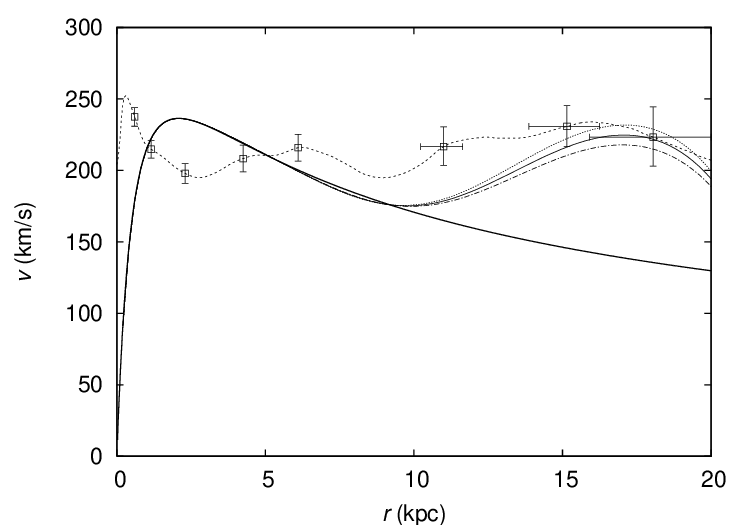}
\caption{High resolution rotation curve of the Milky Way (from (Brownstein \& Moffat 2006)) and the model fit.
  The dashed line with open symbols with error bars are the observational data.
  Thick solid line is the Newtonian galaxy rotation curve that is essentially Eq.(10) with 
  $B_\phi=B_r=0$. Thin solid line is our model best fit that is  Eq.(10) with 
  the magnetic fields specified by Eqs.(7)-(9) and $B_0=const=6.8 $ $\mu$G, $p= 24.12^\circ $, $r_0= 4.54$ kpc.
  In order to show the variation with the magnetic field strength the
  dotted line shows the model with $B_0=6.8\times1.05 $ $\mu$G, and dash dotted line shows the model with 
  $B_0=6.8\times0.95 $ $\mu$G (while
  keeping the same best fit $p=24.12^\circ$ and $r_0=4.54$ kpc)}
\label{label1}
\end{figure}

We gather from Fig.1 that our model, that is the 
Newtonian gravity plus $\vec j \times \vec B$ force (with no non-baryonic dark matter,
MOND or MSTG) provides tolerable fit to the observation rotational curve of the Galaxy only for
$r \geq 15$ kpc but falls short of the observational points in the range $6-12$ kpc. We also show here how the 
model prediction varies with the strength of the magnetic field (dotted and dash-doted curves in Fig.1).
As expected, increase  (by 5 \%) in the magnetic field yields commensurate increase in the rotational velocity, while
decrease  (by 5 \%) in the magnetic field yields decrease in the rotational velocity. The
case of no magnetic field $B_0=0$ recovers the Newtonian rotational curve (thick solid line) which is well below
of the observational data points for all $r$.

\begin{figure}
\includegraphics[width=\columnwidth]{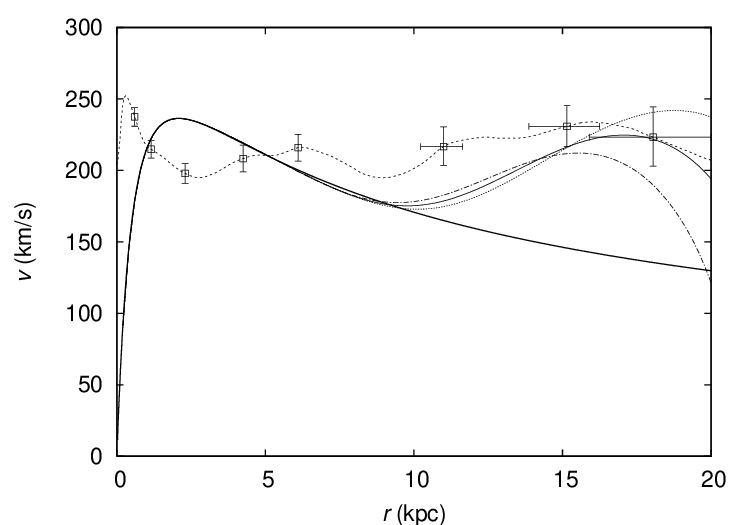}
\caption{The same as in Fig.(1), but here 
  in order to show the variation with the pitch angle  $p$, the
  dotted line shows the model with $p= 24.12\times1.05^\circ $, and dash dotted line shows the model with 
  $p=24.12\times0.95^\circ $ (while
  keeping the same best fit $B_0=6.8 $ $\mu$G and $r_0=4.54$ kpc)}
\label{label2}
\end{figure}

In Fig.2 we study the dependence of the rotational velocity on variation in the pitch angle.
We gather that an increase (by 5\%) in the magnetic field pitch angle $p$ results in
an increase of the rotational velocity for large $r$. While decrease in $p$ produces a lower velocity
beyond $r=15$ kpc. This conclusion is similar to that of Nelson (1988) (also see discussion below).

\begin{figure}
\includegraphics[width=\columnwidth]{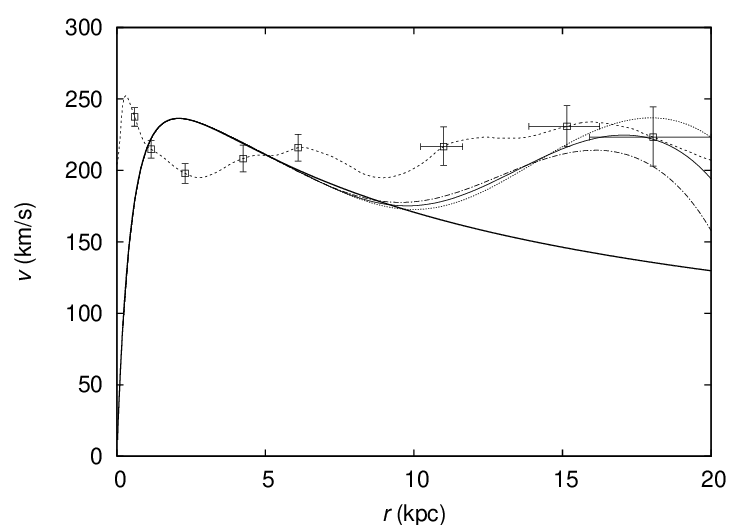}
\caption{The same as in Fig.(1), but here 
  in order to show the variation with the length scale, $r_0$, the
  dotted line shows the model with $r_0= 4.54\times1.05$ kpc, and dash dotted line shows the model with 
  $r_0= 4.54\times0.95$ kpc (while
  keeping the same best fit $B_0=6.8 $ $\mu$G and $p= 24.12^\circ $)}
\label{label3}
\end{figure}

In Fig.3  we investigate how variation in the length scale of the magnetic field $r_0$ affects
the rotational curve. We gather from this graph that an increase in $r_0$ yields an increase
in the rotational velocity, while decrease in $r_0$  produces smaller velocities.
Again, this result seem reasonable because $r_0$ essentially quantifies the extent of the magnetic field.
Hence, as $r_0$  decreases, the effect of $\vec j \times \vec B$ force is weakened and rotation velocity sharply falls down.

In Fig.4 we study whether a better fit than in Figs.(1)--(3) can be achieved by varying the
magnetic field strength
(recall that for the fitting algorithm to converge, in Figs.(1)--(3)
$B_0= 6.8 $ $\mu$G was kept fixed). We thus try to vary $B_0$ by factor of two either way,
i.e. $B_0 \times 2$ and $B_0 /2$. We gather from Fig.4 that the 
small values (e.g. $B_0 /2$) of $B_0$ cannot provide a reasonable fit. However, the large values ($B_0 \times 2$) provide an excellent fit.
Of course, such large values should be discounted on the grounds that such magnetic fields are not observed. However, we remark that starting from about $B_0=11 $ $\mu$G a good fit is possible for all $r$.
This leaves us with a conclusion that in the Galaxy,
whilst only being important for large $r$ ($r \geq 15$ kpc), $\vec j \times \vec B$ force may be important for the gas and plasma dynamics for other galaxies with stronger magnetic fields.

\begin{figure}
\includegraphics[width=\columnwidth]{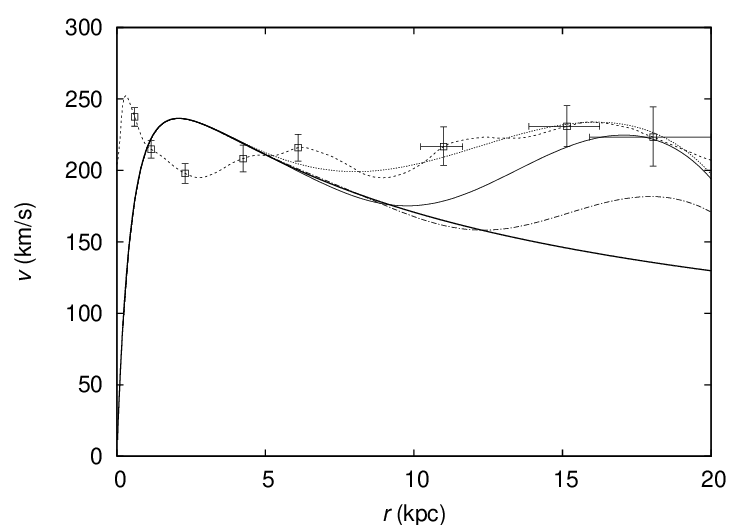}
\caption{The same as in Fig.(1), but here 
  in order to show the variation of the best possible fit with fixed  $B_0$,
   dotted line shows the model with $B_0=6.8\times2=13.6 $ $\mu$G and best fit values  $p= 41.90^\circ$ and $r_0=14.48$ kpc, and dash dotted line shows the model with 
 $B_0=6.8 / 2=3.4 $ $\mu$G and best fit values  $p= 17.74^\circ$ and $r_0=7.43$ kpc.
 Thin solid line is the model best fit with  $B_0=6.8 $ $\mu$G, $p= 24.12^\circ $, $r_0= 4.54$ kpc.}
\label{label4}
\end{figure}

\section{Conclusions}

Earlier work of Nelson (1988) conjectured about importance of the dynamical effect of 
magnetic stress on the tenuous outer gaseous discs of galaxies.
One of the conclusions of
Nelson (1988) was that an increase in the pitch angle of the magnetic field yields higher
rotational velocities (this result is also corroborated in our model -- see our Fig.2).
The model presented here is an improvement on Nelson (1988) work in that we used a more realistic magnetic field
model of Han \& Qiao (1994), and we perform an actual fit to the Milky Way rotational curve. The latter is possible
because our model is simpler and presents an analytical expression for the 
rotational velocity (Eq.(10)) as opposed 
to the need for solving an ordinary differential equation (Eq.(7) from Nelson (1988)).
Other previous developments include  
(Katz 1994; Battaner et al.~1992; Battaner \& Florido 2000; Battaner et al. 2002; 
Battaner \& Florido 2007; Sanchez-Salcedo \& Reyes-Ruiz 2004).

In the present study we investigate the effect of $\vec j \times \vec B$
force on the dynamics of gas and plasma.
We show that as far as rotational curve of gas and plasma of 
the Milky Way is concerned, inclusion of $\vec j \times \vec B$
force only provides a tolerable fit to the 
rotational curve of the Galaxy  for
$r>15$ kpc from the centre, but fails in the intermediate range  $6-12$ kpc.
In principle, a tolerable fit can be obtained for all radii with
the stronger magnetic field of $B_0 \geq 11 $ $\mu$G, but such high values are
not observed.
Further study is needed to clarify whether the model formulated in this work can be used to fit rotational curves of
other known galaxies where the magnetic fields are stronger.

Other weaknesses of this model include:

(i) How well the galactic plasma couples to the magnetic field (for $\vec j \times \vec B$
 to be effective). Naturally this coupling is prescribed by the degree of ionisation of the medium, 
which in turn, is prescribed by the Saha equation and is sensitive to the temperature. 
In general, initial temperatures of galaxies are expected to be high because 
so called virial temperature (page 557 
from (Gilmore et al. 1989))
$T_{\rm virial} \simeq G M m_p /(k R)$, where symbols have
usual meaning, for a typical size galaxy is of the order of
$10^6$ K. However, after cooling phase galactic discs are much cooler at about
 $\simeq 10^4$ K. Quireza et al. (2006) quote electron temperatures in the
disc of galaxy of the order of $10^4$ K which means that degree of ionisation
of the galactic disc is sufficient to couple plasma to the magnetic field
and $\vec j \times \vec B$ force. After all, solar photosphere which is at temperature of
only 6000K is commonly described by MHD approximation, despite the low degree of 
ionisation and the presence of large concentration of neutrals.
Also, in addition to thermal collisions some significant ionisation may be 
provided by the cosmic rays (mostly protons) that are accelerated at the 
bow and termination shocks. A substantial flux of
cosmic rays is produced in a shock at Galactic north, a direction
towards which our Galaxy has long been known to be moving in
the Local Supercluster with the velocity of 200 km s$^{-1}$ (Medvedev \& Melott 2007).

(ii) The origin of the magnetic field 
in the galaxy itself is deeply coupled with the
Galaxy's dynamics and MHD via the dynamo mechanism. 
The field strength and
morphology are dependent on the dynamics of the plasma, which is a 
function of density, temperature, turbulent velocity, and
galactic rotation. 
Therefore, the centrifugal
force due to galactic rotation acts both on the plasma  and
magnetic field, and not on the plasma alone. 

(iii) This work was complete when a study was brought to the author's attention that
provides an improved model for the Galactic magnetic field (van Eck et al. 2011). 
Using observations of the 194 Faraday rotation measures from the Very Large 
Array (VLA), van Eck et al. 2011 have used the following
three models from Sun et al. (2008): a BSS model and two axisymmetric 
spiral (ASS) models, one with magnetic field reversals following the 
spiral arms of the Galaxy (ASS+ARM) and the other with reversals
in rings of constant radius (ASS+RING). Data analysis of van Eck et 
al. 2011 favours an axisymmetric spiral model with reversals
occurring in rings (as opposed to along spiral arms). 
Sun et al. (2008) also conclude that the ASS field configuration plus 
a reversal inside the solar circle either in a ring or between the 
inner edges of the Sagittarius-Carina arm and the Scutum-Crux arm is favoured
by the rotation measures of extragalactic sources observed along 
the Galactic plane. A BSS field configuration, as in Han \& Qiao 1994, 
used in the present study, also fits these rotation measures, 
however, it then fails to fit the observed rotation measures 
gradient in latitude direction. Overall, the conclusion is that the 
axisymmetric spiral models fit the recent (improved) data better. 
However, the use of ASS model instead of BSS model would not change much 
in the area near the Sun's locations
and also between 6 and 12 kpc from the Galactic Center
(this view is also shared by an anonymous referee). Also, the use 
of ASS models (e.g. ASS+RING) would come with a complication that 
the magnetic field is discontinuous (it has reversals (jumps) at certain 
radii), see Eq.(8) from Sun et al. (2008). This means when 
using our Eq.(10) a numerical derivative (e.g. high-order, 
centered finite difference) should be used.
This could be a subject of a future study. 

(iv) Our model best fit, using non-linear least squares Marquardt-Levenberg algorithm,
produces the pitch angle of about 24 degrees. This is larger than the
observed average (median) pitch angle of about 13 degrees. However, the observed angle range
is quite broad 5--25 degrees  (see Table 1 in Vallee 2008).
It would be interesting to test whether the ASS magnetic field models, that are more favoured
by the current observations, could produce more realistic average pitch angles compared to the BSS
model used.

(v) The model gas density, given by Eq.(6) 
and originally used by Brownstein \& Moffat 2006,
does not include the peaks from the density waves nor from nonthermal turbulence, 
 both capable of affecting the rotation curve. Such advanced
topics are naturally beyond the scope of the
simple model presented here.

The overall conclusion of this work is that the $\vec j \times \vec B$ force 
does not play an important role in the Galactic rotational curve plasma dynamics
in the intermediate range of distances $6-12$ kpc from the centre, 
whilst the effect is considerable for larger $r$, $r \geq 15$ kpc, where it is the most crucial.
This may also have more important implications for the rotational curves of galaxies 
where the magnetic fields are stronger than in our Galaxy.

\acknowledgements
Author would like to thank: J.R. Brownstein for providing observational data of Milky Way
rotational curve; 
T. Stanev and J. Alvarez-Muniz for clarifying some 
aspects of the galactic magnetic field model; and an 
anonymous referee whose comments contributed to an improvement of this paper.


\begin{thebibliography}{}

\bibitem{} Alvarez-Muniz, J., Engel, R., Stanev, T.: 2002, ApJ 572, 185
\bibitem{} Battaner, E., Garrido, J.L., Membrado, M., Florido, E.: 1992, Nature 360, 652
\bibitem{} Battaner, E., Florido, E.: 2000, Fundam. of Cosmic Phys. 21,  1
\bibitem{} Battaner, E., Florido, E., Jim\'enez-Vicente, J.: 2002, A\&A 388, 213
\bibitem{} Battaner, E., Florido, E.: 2007, Astron. Nachr. 328, 92
\bibitem{} Brownstein, J.R., Moffat, J.W.: 2006, ApJ 636, 721
\bibitem{} Dobbs, C.L., Price, D.J.: 2008, MNRAS 383, 497
\bibitem{} Gilmore, G., Wyse, R.F.G., Kuijken, K.: 1989, Ann. Rev. Astron. Asrophys. 27, 555
\bibitem{} Gosling, J.T., Pizzo, V.J.: 1999, Space Sci rev. 89, 21
\bibitem{} Han, J.L., Qiao, G.J.: 1994, A\&A 288, 759
\bibitem{} Katz, J.I.: 1994, Ap\&SS 213, 155
\bibitem{} Medvedev, M.V., Melott, A.L.: 2007, ApJ 664, 879
\bibitem{} Milgrom, M.: 1983, ApJ 270, 365
\bibitem{} Moffat, J.W.: 2005, J. Cosmol. Astropart. Phys. 05, 003
\bibitem{} Nelson, A.H.: 1988, MNRAS 233, 115
\bibitem{} Quireza, C., Rood, R.T., Bania, T.M., Balser, D.S., Maciel, W.J.: 2006, ApJ 653,  1226 
\bibitem{} Sanchez-Salcedo, F.J., Reyes-Ruiz, M.: 2004, ApJ 607, 247
\bibitem{} Sofue, Y., Fujimoto, M., Wielebinski, R.: 1986, Ann. Rev. Astron. Asrophys. 24, 459
\bibitem{} Stanev, T.: 1997, ApJ 479, 290
\bibitem{} Sun, X.H., Reich, W., Waelkens, A., Ensslin, T.A.: 2008, A\&A, 477, 573
\bibitem{} Vall\'ee, J.P.: 2004, New Astron. Rev. 48, 763
\bibitem{} Vall\'ee, J.P.: 2008, AJ, 135, 1301
\bibitem{} Van Eck, C.L., Brown, J.C., Stil, J.M., Rae, K., Mao, S.A., Gaensler, B.M., Shukurov, A., Taylor, A.R., 
Haverkorn, M., Kronberg, P.P., McClure-Griffiths, N.M.: 2011, ApJ, 728, 97

\end{thebibliography}
\end{document}